\documentstyle[sprocl,epsf]{article}
\def\be{\begin{equation}}
\def\ee{\end{equation}}
\def\bea{\begin{eqnarray}}
\def\eea{\end{eqnarray}}

\begin{document}
\begin{titlepage}
\hbox{
\vtop{\hbox{}
\hbox to \hsize{\hfill\large hep-ph/9609367}
\vspace{2.mm}
\hbox to \hsize{\hfill\large Fermilab-Conf-96/315-T}
\vspace{2.mm}
\hbox to \hsize{\hfill\large August 1996}}}

\vspace{1.5cm}
\begin{center}
{\LARGE \sc Gluonic Three Jet Production at Next to Leading Order
\footnote{Talk presented at the 1996 Divisional Meeting of the
Division of Particles and Fields of the American Physical Society,
10-15 August, Minneapolis, Minnesota.}\\[1.cm]}
{\large \sc William B. Kilgore \footnote{\tt kilgore@fnal.gov}\\[2.mm]}
{\large \it Fermi National Accelerator Laboratory\\ P.O. Box 500
\\Batavia, IL 60510, USA}

\vspace{1.5cm}
{\large \bf
ABSTRACT \\[6.mm]}
\end{center} 
{ \baselineskip=16pt \large
I report results from a next-to-leading order event generator of
purely gluonic jet production.  This calculation, performed in
collaboration with Walter Giele, is the first step in
the construction of a full next-to-leading order calculation of three
jet production at hadron colliders.}

\vfill

\end{titlepage}

\title{THREE JET PRODUCTION AT NEXT TO LEADING ORDER}
\author{ WILLIAM B. KILGORE }
\address{Fermi National Accelerator Laboratory\\ P.O. Box 500
\\Batavia, IL 60510, USA}
\maketitle\abstracts{
I report results from a next-to-leading order event generator of
purely gluonic jet production.  This calculation, performed in
collaboration with Walter Giele, is the first step in
the construction of a full next-to-leading order calculation of three
jet production at hadron colliders.}
\subsection*{Introduction}
In this talk I will discuss some results of a next-to-leading order
event generator for hadronic three jet production.  I will begin
by briefly outlining the procedure for performing next-to-leading
order jet calculations.  I will then present a status report on the
progress of this calculation.

\subsection*{Next-to-Leading Order Jet Calculations}
The calculation of three jet production at next-to-leading order
combines two-to-three parton scattering to one loop with Born level
two-to-four parton scattering.  Both of these contributions are
singular.  Only the sum of the two contributions is finite and
physically meaningful.  The one loop two-to-three parton amplitudes
contain infrared singularities arising from the presence of nearly
on-shell massless partons in the loops.  The Born
level two-to-four parton amplitudes are also infrared singular,
diverging when one of the partons is very soft or when two partons are
highly collinear.

The origin of the singularities concerns parton resolvability.  If a
parton becomes very soft, or if two partons are highly collinear, it
becomes impossible to resolve all final state partons from one
another.  The four parton final 
state looks instead like a three parton final state.  (In fact, of
course, individual partons can never be resolved, and are only
observed as jets of hadrons.)  By imposing a resolvabilty criterion
one can define an infrared region of phase space.  By using asymptotic
approximations of soft or collinear matrix elements\cite{BGb} within
that region, one can integrate out the unresolved parton and obtain
effective three body matrix elements with poles that exactly cancel
those of the one loop matrix elements.

This is known as the phase space slicing method.\cite{GG,GGK}  The
infrared region is defined by the 
arbitrary resolution parameter $s_{min}$.  If the invariant mass
$s_{ij}$ of two partons labelled $i$ and $j$ is larger than $s_{min}$,
the partons are said to be resolved from one another (although they
may yet be clustered into the same jet), otherwise partons $i$ and $j$
are said to be unresolved from one another. 
If there is only one pair of unresolved partons $i$ and $j$, then
those partons are said to be collinear.  If there 
is some parton $i$ which in unresolvable from two or more partons
$j,k,\dots$, then parton $i$ is said to be soft.

Using phase space slicing, the singularities are removed from the
two-to-four parton scattering process and added to the one-loop
two-to-three process, cancelling the singularities. 
However, since the boundary of the infrared region was defined by the
arbitrary parameter $s_{min}$, the slicing procedure induces
logarithmic $s_{min}$ dependence in both sub-processes.  The
cancellation of the $s_{min}$ dependence in the sum of the two
processes provides an important cross check on the calculation.

The resolution parameter $s_{min}$ is completely arbitrary and is
independent of the jet clustering algorithm.  This 
allows us to use a variety of jet algorithms and
facilitates comparison with experiment.  In principle, $s_{min}$ can
take any value, but in practice must lie within a finite range.  If
$s_{min}$ is too large, it forces partons to be clustered that the
jet algorithm would otherwise leave unclustered.  If $s_{min}$ is too
small, the logarithms of $s_{min}$ become large.  The magnitude of the
cancellation between the two sub-processes grows, requiring increased
computer time to obtain the cancellation to a given statistical
accuracy. 

\subsection*{Progress Report}
At this time, we have developed a working event generator for pure
gluon scattering.  We thus combine the one-loop virtual cross section
for $gg \rightarrow ggg$ scattering\cite{BDKa} with the Born level
cross section for $gg \rightarrow gggg$.  This development is a
significant step towards completing an event generator for the full
three jet cross section at next to leading order, since all essential
components such as phase space generators, jet clustering algorithms,
phase space slicing, etc., must be working properly.  In principle,
the completion of the project simply involves adding the remaining matrix
elements to the existing program structure.

\begin{figure}[t]
\epsfxsize=\hsize\epsfbox{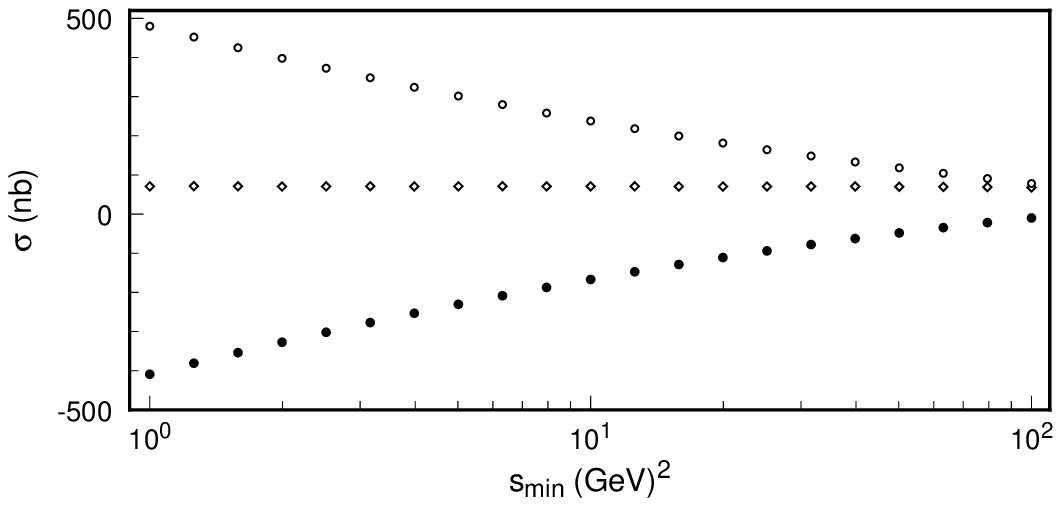}
\epsfxsize=\hsize\epsfbox{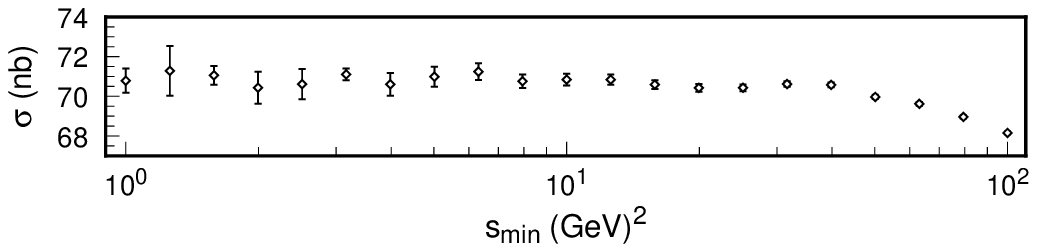}
\caption{a) $s_{min}$ dependence of the total cross section
(center), $gg\rightarrow gggg$ sub-process (top) and $gg\rightarrow
ggg$ sub-process (bottom). b) Expanded view of the $s_{min}$
dependence of the total cross section.}
\label{fig:smin}
\end{figure}

In Figure~\ref{fig:smin}, I show the $s_{min}$ dependence of the total
cross section and of the two component subprocesses.  Clearly, the
calculation is well behaved for any value of $s_{min}$ below
approximately $30$ GeV${}^2$.  Above that value, the resolution
parameter interferes with the jet algorithm and forces excessive
parton clustering.  As expected, good statistical accuracy is more
difficult to obtain at small values of $s_{min}$.

\section*{Acknowledgments} Fermilab is operated by Universities
Research Association, Inc., under contract DE-AC02-76CH03000 with the
U.S. Department of Energy.

\end{document}